\documentclass[12pt]{article}
\usepackage[left=2.5cm,top=2.50cm,right=2.5cm,bottom=2.50cm]{geometry}
\usepackage{mathrsfs}
\usepackage{amsmath,amssymb,latexsym,color,cancel,graphicx,bbm,colortbl}
\usepackage[english]{babel}
\usepackage[latin1]{inputenc}
\usepackage{ragged2e}
\usepackage{cite}
\begin{document}
\date{}

\title{Matrix diagonalization and exact solution of the k-photon Jaynes-Cummings model}
\author{E. Chore$\tilde{n}$o$^{a}$, \footnote{{\it E-mail address:} echorenoo0800@alumno.ipn.mx}\\ D. Ojeda-Guill\'en$^{b}$ and V.D. Granados$^{a}$} \maketitle

\begin{minipage}{0.9\textwidth}
\small $^{a}$ Escuela Superior de F{\'i}sica y Matem\'aticas,
Instituto Polit\'ecnico Nacional, Ed. 9, Unidad Profesional Adolfo L\'opez Mateos, Delegaci\'on Gustavo A. Madero, C.P. 07738, Ciudad de M\'exico, Mexico.\\
\small $^{b}$ Escuela Superior de C\'omputo, Instituto Polit\'ecnico Nacional,
Av. Juan de Dios B\'atiz esq. Av. Miguel Oth\'on de Mendiz\'abal, Col. Lindavista,
Delegaci\'on Gustavo A. Madero, C.P. 07738, Ciudad de M\'exico, Mexico.\\

\end{minipage}

\begin{abstract}

We study and exactly solve the two-photon and k-photon Jaynes-Cummings models by using a novelty algebraic method. This algebraic method is based on the Pauli matrices realization and the tilting transformation of the $SU(2)$ group and let us diagonalize the Hamiltonian of these models by properly choosing the coherent state parameters of the transformation. Finally, we explicitly obtain the energy spectrum and eigenfunctions for each model.\\

\end{abstract}


\section{Introduction}
In the study of quantum optics models, one of the fundamental theoretical paradigms is the Jaynes-Cummings model \cite{Jay,Dod,shore}. This is the simplest and completely soluble quantum-mechanical model and describes the interaction between a two-level atom and a quantized field. The exact solution of this model has been found by using the rotating wave approximation (RWA) \cite{Haroche}. Under this approximation, the Hamiltonian terms that do not conserve the number of excitations in the system are neglected, since this terms oscillate rapidly and their contributions are very small except for very high intensity fields. These solutions within this approximation yield quantum collapse and revival of atomic inversion \cite{Narozhny}, squeezing of the radiation field \cite{Kuklinski}, among other quantum effects. All these effects have been corroborated experimentally, as can be seen in references \cite{Goy,Brune,Guerlin}. Moreover, the Dirac-Moshinsky oscillator has been connected with the Jaynes-Cummings model in $1+1$ and $2+1$ dimensions by using the usual creation and annihilation operators, as can be seen in references \cite{Toyama,Nogami,Rados,Nos,Bermudez,Chiral}.

The Jaynes-Cummings model has also been the subject of many generalizations \cite{Koch,Buze,Gou,JCAJC}, besides some of its generalization are still under study as shown in references \cite{Lam,Ret,Kop,Sun}. To study some of these generalizations several algebraic methods have been used, such as the q-analog of the Holstein-Primakoff realization of the $su(1,1)$ Lie algebra to give solution of the Deformed Jaynes-Cummings model \cite{Cha}.

The two-photon Jaynes-Cummings model was introduced to study the coupling between a single atom and a single-mode cavity field, with the atom making two-photon transitions \cite{Buck80}. This model has attracted a lot of attention and has been studied interacting with the squeezed vacuum \cite{Gerry}, in a Kerr-like medium \cite{Joshi}, among other works \cite{Bartzis,Zhou,Tahira,Ng}. The multiphoton generalization of this model was introduced shortly after the two-photon model by Sukumar and Buck to study the atomic dynamics of such models interacting with coherent light \cite{Sukumar81}. In these works, the authors showed that these models exhibit periodic decay and revival of atomic coherence. The k-photon Jaynes-Cummings model has been extensively studied, as can be seen in references \cite{Shumovsky,Vogel,Baghshahi,kocr,Pana}.

The aim of the present work is to obtain the energy spectrum and eigenfunctions of the two-photon and k-photon Jaynes-Cummings models by using the tilting transformation and a realization of the $su(2)$ Lie algebra in terms of the Pauli matrices.

This work is organized as follows. In Section 2, we study the two-photon Jaynes-Cummings model by means of the $SU(2)$ tilting transformation. In this approach, we write the Hamiltonian in the interaction picture in terms of the Pauli matrices realization of the $su(2)$ Lie algebra. By choosing properly the coherent state parameters of the transformation, we diagonalize the interaction Hamiltonian and obtain the energy spectrum of this model. The eigenfunctions are obtained by introducing a matrix form of the $SU(2)$ displacement operator. In Section 3, we proceed as in the previous Section to study the k-photon Jaynes-Cummings model. Finally, we give some concluding remarks.

\section{Exact solution of the two-photon Jaynes-Cummings model}

The Hamiltonian of two-photon Jaynes-Cummings model with rotating wave approximation (RWA) is given by \cite{kocr,Pana}
\begin{equation}
\hat{H}=\hbar\omega\hat{a}^{\dag}\hat{a}+\frac{\hbar\omega_{0}}{2}\sigma_{0}+\lambda\left(\sigma_{+}(\hat{a})^{2}+\sigma_{-}(\hat{a}^{\dag})^{2}\right)\label{HP},
\end{equation}
where $\lambda$ is the atom-cavity coupling constant, $\omega$ and $\omega_{0}$ are the frequencies of the atom and cavity respectively, $\hat{a}$ and $\hat{a}^{\dag}$ are the bosonic annihilation and creation operators that obey the commutating relation $[\hat{a},\hat{a}^{\dag}]=1$. $\sigma_{0}$ and $\sigma_{\pm}$ are the Pauli matrices of the two-level atom, which can be used to construct a realization of the $su(2)$ Lie algebra (see equation (\ref{real}) of Appendix). With this realization, and adding and subtracting the term $2\hbar\omega J_{0}$, the Hamiltonian of two-photon Jaynes-Cummings can be expressed as
\begin{equation}
H=\hbar\omega(\hat{a}^{\dag}\hat{a}+2J_{0})+\left(\hbar\omega_{0}-2\hbar\omega\right)J_{0}+\lambda\left(J_{+}(\hat{a})^{2}+J_{-}(\hat{a}^{+})^{2}\right).
\end{equation}
Since the term $\hat{a}^{\dag}\hat{a}+2J_{0}$ is diagonal and commutes with each of the other terms of the Hamiltonian, we can work only with the interaction Hamiltonian $H_I$
\begin{equation}
\hat{H}_{I}=\left(\hbar\omega_{0}-2\hbar\omega\right)J_{0}+\lambda\left(J_{+}(\hat{a})^{2}+J_{-}(\hat{a}^{+})^{2}\right).
\end{equation}
By means of the $SU(2)$ displacement operator $D(\xi)=\exp(\xi J_{+}-\xi^{*}J_{-})$ (see Appendix), we apply the tilting transformation to the Schr\"odinger equation  $\hat{H}_{I}\Psi=E_{I}\Psi$ to obtain \cite{JCAJC,gerryberry,Nos1,Nos2}
\begin{equation*}
D^{\dag}(\xi)\hat{H_{I}}D(\xi)D^{\dag}(\xi)\Psi =E_{I}D^{\dag}(\xi)\Psi,
\end{equation*}
\begin{equation}
\hat{H}_{I}'\Psi'=E_{I}\Psi'\label{TilEc}.
\end{equation}
Notice that in these expressions $H_{I}'=D^{\dag}(\xi)H_{I}D(\xi)$ is the tilted interaction Hamiltonian and $\Psi'$ its wave function. Thus, by using equations (\ref{DJ+}), (\ref{DJ-}) and (\ref{DJ0}) of Appendix, the tilted Hamiltonian can be written as
\begin{align}
\hat{H}_{I}'=&\left[(\hbar\omega_{0}-2\hbar\omega)\left(2\epsilon + 1 \right)-\frac{\lambda\xi^{*}\delta}{|\xi|}(\hat{a})^{2} -\frac{\lambda\xi\delta}{|\xi|}(\hat{a}^{\dag})^{2}\right]J_{0}\nonumber\\&+\left[(\hbar\omega_{0}-2\hbar\omega)\frac{\delta\xi}{2|\xi|}+\lambda(\epsilon+1)(\hat{a})^{2}+\frac{\lambda\epsilon\xi}{\xi^{*}}(\hat{a}^{\dag})^{2}\right]J_{+}\nonumber\\&+\left[(\hbar\omega_{0}-2\hbar\omega)\frac{\delta\xi^{*}}{2|\xi|}+\frac{\lambda\epsilon\xi^{*}}{\xi}(\hat{a})^{2}+\lambda(\epsilon+1)(\hat{a}^{\dag})^{2}\right]J_{-}\label{tilTPC}.
\end{align}
Therefore, the tilted interaction Hamiltonian $H_{I}'$ can be written in the following matrix form
\begin{equation}
\hat{H}_{I}'=\begin{pmatrix}(\frac{\hbar\omega_{0}}{2}-\hbar\omega)(2\epsilon + 1)-\frac{\lambda\xi^{*}\delta(\hat{a})^{2}}{2|\xi|} -\frac{\lambda\xi\delta(\hat{a}^{\dag})^{2}}{2|\xi|}&\frac{(\hbar\omega_{0}-2\hbar\omega)\delta\xi}{2|\xi|}+\lambda(\epsilon+1)(\hat{a})^{2}+\frac{\lambda\epsilon\xi(\hat{a}^{\dag})^{2}}{\xi^{*}}\\ \frac{(\hbar\omega_{0}-2\hbar\omega)\delta\xi^{*}}{2|\xi|}+\frac{\lambda\epsilon\xi^{*}(\hat{a})^{2}}{\xi}+\lambda(\epsilon+1)(\hat{a}^{\dag})^{2}&(\hbar\omega-\frac{\hbar\omega_{0}}{2})(2\epsilon + 1)+\frac{\lambda\xi^{*}\delta(\hat{a})^{2}}{2|\xi|} +\frac{\lambda\xi\delta(\hat{a}^{\dag})^{2}}{2|\xi|}\\\end{pmatrix}.\label{Matrix}
\end{equation}

In order to find the eigenvalues of this matrix Hamiltonian, we assume that the eigenvectors of $H_{I}'$ are of the form
\begin{equation}
{\phi'}^{(1)}_{n}=\begin{pmatrix}|{\psi'}^{(1)}_{n}\rangle\\0\end{pmatrix}, \quad\quad
{\phi'}^{(2)}_{n'}=\begin{pmatrix}0\\|{\psi'}^{(2)}_{n'}\rangle\end{pmatrix}\label{vectors},
\end{equation}
where $|{\psi'}^{(1)}_{n}\rangle$ and $|{\psi'}^{(2)}_{n'}\rangle$ are the states of the one-dimensional harmonic oscillator, ${E}^{(1)}_{I}$ and ${E}^{(2)}_{I}$ are their corresponding eigenvalues, with ${E}^{(1)}_{I}\neq {E}^{(2)}_{I}$. For the eigenvector ${\phi'}^{(1)}_{n}$ we find from the eigenvalues equation $H_{I}'{\phi'}^{(1)}_{n}={E}^{(1)}_{I}{\phi'}^{(1)}_{n}$
\begin{equation}
\left[\left(\frac{\hbar\omega_{0}}{2}-\hbar\omega\right)\left(2\epsilon + 1 \right)-\frac{\lambda\xi^{*}\delta}{2|\xi|}(\hat{a})^{2} -\frac{\lambda\xi\delta}{|2\xi|}(\hat{a}^{\dag})^{2}\right]|{\psi'_{n}}^{(1)}\rangle={E}^{(1)}_{I}|{\psi'}^{(1)}_{n}\rangle\label{E1}
\end{equation}
\begin{equation}
\left[\left(\hbar\omega_{0}-2\hbar\omega\right)\frac{\delta\xi^{*}}{2|\xi|}+\frac{\lambda\epsilon\xi^{*}}{\xi}(\hat{a})^{2}+\lambda(\epsilon+1)(\hat{a}^{\dag})^{2}\right]|{\psi'_{n}}^{(1)}\rangle=0.\label{cero}
\end{equation}
The equation (\ref{cero}) is satisfied by choosing the coherent state parameters $\theta$ and $\varphi$ as
\begin{equation}
\theta=\tan^{-1}\left(\frac{2\lambda\sqrt{\hat{a}^{2}(\hat{a}^{\dag})^{2}}}{\hbar\omega_{0}-2\hbar\omega}\right),\quad\quad\varphi=i\ln{\left[\frac{(\hbar\omega_{0}-2\hbar\omega)\delta}{\lambda(2\epsilon+1)(\hat{a}^{\dag})^{2}} \right]}\label{Parameters}.
\end{equation}
By substituting these parameters into the matrix Hamiltonian of equation (\ref{Matrix}), we obtain for ${\phi'}^{(1)}_{n}$ the following expression
\begin{equation}
\frac{1}{2}\begin{pmatrix}\pm\sqrt{(\hbar\omega_{0}-2\hbar\omega)^{2}+4\lambda\hat{a}^{2}\hat{a}^{\dag}{}^{2}}&0\\ 0&\mp\sqrt{(\hbar\omega_{0}-2\hbar\omega)^{2}+4\lambda\hat{a}^{2}\hat{a}^{\dag}{}^{2}}\\\end{pmatrix}\begin{pmatrix}|{\psi'}^{(1)}_{n}\\0\end{pmatrix}=\hat{E}^{(1)}_{I}\begin{pmatrix}|{\psi'}^{(1)}_{n}\rangle\\0\end{pmatrix}\label{diagonal}.
\end{equation}
Hence, the coherent state parameters of equation (\ref{Parameters}) diagonalize the interaction Hamiltonian $H_{I}'$ and the energy ${E}^{(1)}_{I}$ for the spinor component $|{\psi'}^{(1)}_{n}\rangle$ is given by
\begin{equation}
\hat{E}^{(1)}_{I}|{\psi'}^{(1)}_{n}\rangle=\pm\frac{1}{2}\sqrt{(\hbar\omega_{0}-2\hbar\omega)^{2}+4\lambda\hat{a}^{2}(\hat{a}^{\dag})^{2}}|{\psi'}^{(1)}_{n}\rangle.
\end{equation}
Therefore, since the state $|{\psi'}^{(1)}_{n}\rangle$ is a number state of one-dimensional harmonic oscillator, the energy spectrum ${E}^{(1)}_{I}(n)$ is found to be
\begin{equation}
{E}^{(1)}_{I}(n)=\pm\frac{1}{2}\sqrt{(\hbar\omega_{0}-2\hbar\omega)^{2}+4\lambda(n+1)(n+2)}\label{En1}.
\end{equation}
It is necessary to mention that for this energy spectrum we have chosen a particular order for the multiplication of the creation and annihilation operators, namely, anti-normal ordering. Also, notice that the operator $\frac{1}{(\hat{a}^{\dag})^{2}}$ is the inverse operator of $(\hat{a}^{\dag})^{2}$ in this ordering i.e.,
\begin{equation}
 (\hat{a}^{\dag})^{2}\frac{1}{(\hat{a}^{\dag})^{2}}=1.
 \end{equation}
Analogously, if we apply the same procedure to the eigenvector ${\phi'}^{(2)}_{n'}$ we find the equations
\begin{equation}
\left[\left(\hbar\omega-\frac{\hbar\omega_{0}}{2}\right)(2\epsilon + 1)+\frac{\lambda\xi^{*}\delta(\hat{a})^{2}}{2|\xi|} +\frac{\lambda\xi\delta(\hat{a}^{\dag})^{2}}{2|\xi|}\right]|{\psi'}^{(2)}_{n'}\rangle={E}^{(2)}_{I}|{\psi'}^{(2)}_{n'}\rangle,\label{eige,oper}
\end{equation}
\begin{equation}
\left[\frac{(\hbar\omega_{0}-2\hbar\omega)\delta\xi}{2|\xi|}+\lambda(\epsilon+1)(\hat{a})^{2}+\frac{\lambda\epsilon\xi(\hat{a}^{\dag})^{2}}{\xi^{*}}\right]|{\psi'}^{(2)}_{n'}\rangle=0,\label{eige,oper2}
\end{equation}
where equation (\ref{eige,oper2}) is again satisfied by choosing the same values of the coherent state parameters $\theta$ and $\varphi$ given in (\ref{Parameters}) and if $n'=n$, as it is shown in equation (\ref{diagonal}). However, these equations are also satisfied by choosing the value of these coherent state parameters as
\begin{equation}
\theta'=\tan^{-1}\left(\frac{2\lambda\sqrt{(\hat{a}^{\dag})^{2}\hat{a}^{2}}}{\hbar\omega_{0}-2\hbar\omega}\right),\quad\quad\varphi=i\ln{\left[\frac{(\hbar\omega_{0}-2\hbar\omega)\delta}{\lambda(2\epsilon+1)(\hat{a}^{\dag})^{2}} \right]}\label{Parameters2},
\end{equation}
where the only difference between both solutions is that the product of the bosonic annihilation and creation operators is exchanged in the value of the parameter $\theta$. Therefore, since in principle ${E}^{(1)}_{I}\neq {E}^{(2)}_{I}$, we have chosen a normal ordering for the multiplication of the creation and annihilation operators, and $\frac{1}{(\hat{a}^{\dag})^{2}}$ as a operator such that
\begin{equation}
\frac{1}{(\hat{a}^{\dag})^{2}}(\hat{a}^{\dag})^{2}=1-|0\rangle\langle{0}|-|1\rangle\langle1|.\label{in2}
\end{equation}
By substituting of the coherent state parameters (\ref{Parameters2}) into the matrix expression of the Hamiltonian (\ref{Matrix}), the interaction Hamiltonian $\hat{H}_{I}'$ becomes a diagonal matrix given by
\begin{equation}
\hat{H}_{I}'=\begin{pmatrix}\pm\frac{1}{2}\sqrt{(\hbar\omega_{0}-2\hbar\omega)^{2}+4\lambda\hat{a}^{\dag}{}^{2}\hat{a}^{2}}&0\\ 0&\mp\frac{1}{2}\sqrt{(\hbar\omega_{0}-2\hbar\omega)^{2}+4\lambda\hat{a}^{\dag}{}^{2}\hat{a}^{2}}\\\end{pmatrix}.
\end{equation}
Therefore, from the equation of eigenvalues $H_{I}'{\phi'}^{(2)}_{n'}={E}^{(1)}_{I}{\phi'}^{(2)}_{n'}$ we have the expression
\begin{equation}
\hat{E}^{(2)}_{I}|{\psi'}^{(2)}_{n'}\rangle=\mp\frac{1}{2}\sqrt{(\hbar\omega_{0}-2\hbar\omega)^{2}+4\lambda(\hat{a}^{\dag})^{2}\hat{a}^{2}}|{\psi'}^{(2)}_{n'}\rangle.\label{ene2}
\end{equation}
By using relationships (\ref{in2}) and (\ref{ene2}), we find that the eigenvalue ${E}^{(2)}_{I}$ is well defined only for integer numbers in range of $n'\geq2$ i.e.,
\begin{equation}
{E}^{(2)}_{I}(n')=\mp\frac{1}{2}\sqrt{(\hbar\omega_{0}-2\hbar\omega)^{2}+4\lambda n'(n'-1)}\label{En2}, \quad\quad n'\geq2.
\end{equation}
Thus, we have shown that ${\phi'}^{(1)}_{n}$ and ${\phi'}^{(2)}_{n'}$ are eigenvectors of the matrix Hamiltonian $H_{I}'$ if we choose the coherent state parameters as in equations (\ref{Parameters}) and (\ref{Parameters2}), with its respective commutation order of the bosonic annihilation and creation operators. Moreover, we have obtained the eigenvalues $E^{(1)}_{I}$ and $E^{(2)}_{I}$ for each of these eigenvectors.

On the other hand, if we impose that both eigenvalues $E^{(1)}_{I}$ and $E^{(2)}_{I}$ of the eigenvectors ${\phi'}^{(1)}_{n}$ and ${\phi'}^{(2)}_{n'}$ belong to the same solution, we can build the spinor $\Psi'$ which is solution of equation (\ref{TilEc}). This leads to the fact that the spinor components $|{\psi'}^{(1)}_{n}\rangle$ and $|{\psi'}^{(2)}_{n'}\rangle$ satisfy the relationship $n'\Rightarrow n+2$. Therefore, the spinor $\Psi'$ for the matrix Hamiltonian $H'_{I}$ is
\begin{equation}
\Psi'_{n}=\begin{pmatrix}
|{\psi'}^{(1)}_{n}\rangle\\
|{\psi'}^{(2)}_{n+2}\rangle
\end{pmatrix},\label{estado2}  \quad\quad n=0, 1, 2, ...
\end{equation}
where the states $|{\psi'}^{(1)}_{n}\rangle$ and $|{\psi'}^{(2)}_{n}\rangle$ are the number states of the one-dimensional harmonic oscillator.

The eigenfunctions $\Psi_{n}$ of the two-photon Jaynes-Cummings model are obtained by applying the displacement operator $D(\xi)$ to the spinor $\Psi'_{n}$, $\Psi_{n}=D(\xi)\Psi'_{n}$. In order to do this, we can write the displacement operator $D(\xi)=\exp(\xi J_{+}-\xi^{*}J_{-})$ in the following matrix form
\begin{equation}
D(\xi)=\exp(\xi\sigma_{+}-\xi^{*}\sigma_{-})=\exp\left[\begin{pmatrix}0&\xi\\-\xi^{*}&0\\
\end{pmatrix}\right].
\end{equation}
By expanding the exponential in Taylor series, the even and odd powers of the matrix
\begin{equation}
A=\begin{pmatrix}0&\xi\\-\xi^{*}&0\\
\end{pmatrix}
\end{equation}
may be written as
\begin{equation}
A^{2k+1}=(-1)^{k}|\xi|^{2k}\begin{pmatrix}0&\xi\\-\xi^{*}&0\\\end{pmatrix},\quad\quad A^{2k}=(-1)^{k}|\xi|^{2k}\begin{pmatrix}1&0\\0&1\\\end{pmatrix}.
\end{equation}
Thus, the displacement operator $D(\xi)$ is given by
\begin{equation}
D(\xi)=\sum_{k=0}^{\infty}\frac{(-1)^{k}|\xi|^{2k}}{(2k+1)!}\begin{pmatrix}0&\xi\\-\xi^{*}&0\\\end{pmatrix}+\sum_{k=0}^{\infty}\frac{(-1)^{k}|\xi|^{2k}}{(2k)!}\begin{pmatrix}1&0\\0&1\\\end{pmatrix},
\end{equation}
which can be expressed in terms of the Taylor series for $\sin(x)$ and $\cos(x)$ as
\begin{equation}
D(\xi)=\begin{pmatrix}\cos(|\xi|)&\frac{\xi}{|\xi|}\sin(|\xi|)\\-\frac{\xi^{*}}{|\xi|}\sin(|\xi|)&\cos(|\xi|)\\\end{pmatrix}.
\end{equation}
Now, by using the coherent state parameters (\ref{Parameters}) it is easy to find the following relationships
\begin{equation}
\cos(|\xi|)=\frac{1}{\sqrt{2}}\sqrt{1+\frac{\Delta}{E_{I}}},\quad\quad\sin(|\xi|)=\frac{1}{\sqrt{2}}\sqrt{1-\frac{\Delta}{E_{I}}},
\end{equation}
\begin{equation}
\frac{\xi}{|\xi|}=-\frac{\sqrt{(n+1)(n+2)}}{\hat{a}^{\dag}{}^{2}},\quad\quad\frac{\xi^{*}}{|\xi|}=-\frac{\hat{a}^{\dag}{}^{2}}{\sqrt{(n+1)(n+2)}}.
\end{equation}
Therefore, the $SU(2)$ displacement operator $D(\xi)$ can be expressed in matrix form as
\begin{equation}
D(\xi)=\begin{pmatrix}\frac{1}{\sqrt{2}}\sqrt{1+\frac{\Delta}{E_{I}}}&-\frac{1}{\sqrt{2}}\sqrt{1-\frac{\Delta}{E_{I}}}\quad\frac{\sqrt{(n+1)(n+2)}}{(\hat{a}^{\dag})^{2}}\\\frac{1}{\sqrt{2}}\sqrt{1-\frac{\Delta}{E_{I}}}\quad\frac{(\hat{a}^{\dag})^{2}}{\sqrt{(n+1)(n+2)}} & \frac{1}{\sqrt{2}}\sqrt{1+\frac{\Delta}{E_{I}}}\\\end{pmatrix},\label{Mat}
\end{equation}
where $\Delta=\hbar\omega_{0}-2\hbar\omega$ and $E_{I}$ is the energy eigenvalue of the spinor (\ref{estado2})
\begin{equation}
E_{I}(n)=\pm\frac{1}{2}\sqrt{\Delta^{2}+4\lambda(n+1)(n+2)}.
\end{equation}
It is easy to see that the action of the matrix displacement operator $D(\xi)$ on $\Psi'_{n}$ transforms its spinor components to
\begin{equation}
|{\psi}^{(1)}_{n}\rangle=\frac{1}{\sqrt{2}}\left(\sqrt{1+\frac{\Delta}{E_{I}}}-\sqrt{1-\frac{\Delta}{E_{I}}}\right)|{\psi'}^{(1)}_{n}\rangle,
\end{equation}
\begin{equation}
|{\psi}^{(2)}_{n+2}\rangle=\frac{1}{\sqrt{2}}\left(\sqrt{1+\frac{\Delta}{E_{I}}}+\sqrt{1-\frac{\Delta}{E_{I}}}\right)|{\psi'}^{(2)}_{n+2}\rangle.
\end{equation}
Hence, the normalized spinor of the Hamiltonian of the two-photon Jaynes-Cummings model given by equation (\ref{HP}) is
\begin{equation}
\Psi_{n}=\begin{pmatrix}
\frac{\sqrt{2}E_{I}}{\sqrt{E_{I}+\Delta}-\sqrt{E_{I}-\Delta}}|{\psi}^{(1)}_{n}\rangle\\\frac{\sqrt{2}E_{I}}{\sqrt{E_{I}+\Delta}+\sqrt{E_{I}-\Delta}}|{\psi}^{(2)}_{n+2}\rangle
\end{pmatrix},\label{estado3}  \quad\quad n=0, 1, 2, ...
\end{equation}
with its corresponding energy eigenvalues
\begin{equation}
E_{n}=\hbar\omega(n+1)\pm\frac{1}{2}\sqrt{(\hbar\omega_{0}-2\hbar\omega)^{2}+4\lambda(n^{2}+3n+2)},\label{energy} \quad\quad n=0, 1, 2, ...
\end{equation}
Therefore, we have obtained the eigenfunctions and energy spectrum of the two-photon Jaynes-Cummings model by using the tilting transformation and the $SU(2)$ realization of the Pauli matrices. The energy spectrum of equation (\ref{energy}) coincides with the previously obtained in reference \cite{Pana} by means of the Bargmann-Segal representation.

\section{Exact solution of the k-photon Jaynes-Cummings model.}

In this Section, we shall use the same formalism developed in Section $2$ to obtain the general solution of the k-photon Jaynes-Cummings model. In the rotating wave approximation (RWA), the k-photon Jaynes-Cummings model is given by the expression \cite{kocr,Pana}
\begin{equation}
\hat{H}=\hbar\omega\hat{a}^{\dag}\hat{a}+\frac{\hbar\omega_{0}}{2}\sigma_{0}+\lambda\left(\sigma_{+}(\hat{a})^{k}+\sigma_{-}(\hat{a}^{+})^{k}\right).\label{KPH}
\end{equation}
Again, for convenience we shall focus on the Hamiltonian in the interaction picture
\begin{equation}
\hat{H}_{I}=\left(\hbar\omega_{0}-k\hbar\omega\right)J_{0}+\lambda\left(J_{+}(\hat{a})^{k}+J_{-}(\hat{a}^{+})^{k}\right),
\end{equation}
where we have used the $SU(2)$ realization of the Pauli matrices of equation (\ref{real}).
Now, we apply the tilting transformation to the eigenvalue equation $\hat{H}_{I}\Psi=E_{I}\Psi$ in order to remove the ladder operators $J_{\pm}$
\begin{equation}
D^{\dagger}(\xi)\left[\left(\hbar\omega_{0}-k\hbar\omega\right)J_{0}+\lambda\left(J_{+}(\hat{a})^{k}+J_{-}(\hat{a}^{\dag})^{k}\right)\right]D(\xi)D^{\dagger}(\xi)\Psi=E_{I}D^{\dagger}(\xi)\Psi,
\end{equation}
where again $D(\xi)$ is the $SU(2)$ displacement operator and $\xi=-\frac{1}{2}\theta e^{-i\varphi}$ (see Appendix). If we define the tilted Hamiltonian $\hat{H}_{I}'=D^{\dagger}(\xi)\hat{H}_{I}D(\xi)$ and the wave function $\Psi'=D^{\dagger}(\xi)\Psi$, this equation can be written as $\hat{H}'_{I}\Psi'=E_{I}\Psi'$. Therefore, from equations (\ref{DJ+}), (\ref{DJ-}) and (\ref{DJ0}) of Appendix we find that the tilted Hamiltonian in the interaction picture results to be
\begin{align}
\hat{H}_{I}'=&\left[(\hbar\omega_{0}-k\hbar\omega)\left(2\epsilon + 1 \right)-\frac{\lambda\xi^{*}\delta}{|\xi|}(\hat{a})^{k} -\frac{\lambda\xi\delta}{|\xi|}(\hat{a}^{\dag})^{k}\right]J_{0}\nonumber\\&+\left[(\hbar\omega_{0}-k\hbar\omega)\frac{\delta\xi}{2|\xi|}+\lambda(\epsilon+1)(\hat{a})^{k}+\frac{\lambda\epsilon\xi}{\xi^{*}}(\hat{a}^{\dag})^{k}\right]J_{+}\nonumber\\&+\left[(\hbar\omega_{0}-k\hbar\omega)\frac{\delta\xi^{*}}{2|\xi|}+\frac{\lambda\epsilon\xi^{*}}{\xi}(\hat{a})^{k}+\lambda(\epsilon+1)(\hat{a}^{\dag})^{k}\right]J_{-}\label{tilTPC2}.
\end{align}
In this manner, the tilted interaction Hamiltonian can be written in the following matrix form
\begin{equation}
\hat{H}_{I}'=\begin{pmatrix}\frac{(\hbar\omega_{0}-k\hbar\omega)}{2}(2\epsilon + 1)-\frac{\lambda\xi^{*}\delta(\hat{a})^{k}}{2|\xi|} -\frac{\lambda\xi\delta(\hat{a}^{\dag})^{k}}{2|\xi|}&\frac{(\hbar\omega_{0}-k\hbar\omega)\delta\xi}{2|\xi|}+\lambda(\epsilon+1)(\hat{a})^{k}+\frac{\lambda\epsilon\xi(\hat{a}^{\dag})^{k}}{\xi^{*}}\\ \frac{(\hbar\omega_{0}-k\hbar\omega)\delta\xi^{*}}{2|\xi|}+\frac{\lambda\epsilon\xi^{*}(\hat{a})^{k}}{\xi}+\lambda(\epsilon+1)(\hat{a}^{\dag})^{k}&\frac{(k\hbar\omega-\hbar\omega_{0})}{2}(2\epsilon + 1)+\frac{\lambda\xi^{*}\delta(\hat{a})^{k}}{2|\xi|} +\frac{\lambda\xi\delta(\hat{a}^{\dag})^{k}}{2|\xi|}\\\end{pmatrix}.
\end{equation}

If we consider the vectors ${\phi'}^{(1)}_{n}$ and ${\phi'}^{(2)}_{n}$ of equation (\ref{vectors}) as eigenfunctions of this matrix, we obtain for ${\phi'}^{(1)}_{n}$ the following equations
\begin{equation}
\left[\left(\frac{\hbar\omega_{0}-k\hbar\omega}{2}\right)\left(2\epsilon + 1 \right)-\frac{\lambda\xi^{*}\delta}{2|\xi|}(\hat{a})^{k} -\frac{\lambda\xi\delta}{|2\xi|}(\hat{a}^{\dag})^{k}\right]|{\psi'_{n}}^{(1)}\rangle={E}^{(1)}_{I}|{\psi'}^{(1)}_{n}\rangle\label{EK1}
\end{equation}
\begin{equation}
\left[\frac{(\hbar\omega_{0}-k\hbar\omega)\delta\xi^{*}}{2|\xi|}+\frac{\lambda\epsilon\xi^{*}}{\xi}(\hat{a})^{k}+\lambda(\epsilon+1)(\hat{a}^{\dag})^{k}\right]|{\psi'_{n}}^{(1)}\rangle=0,
\end{equation}
where $E^{(1)}_{I}$ is the corresponding eigenvalue. If we choose the coherent state parameters $\theta$ and $\varphi$ as
\begin{equation}
\theta=\tan^{-1}\left(\frac{2\lambda\sqrt{\hat{a}^{k}(\hat{a}^{\dag})^{k}}}{\hbar\omega_{0}-k\hbar\omega}\right),\quad\quad\varphi=i\ln{\left[\frac{(\hbar\omega_{0}-k\hbar\omega)\delta}{\lambda(2\epsilon+1)(\hat{a}^{\dag})^{k}} \right]}\label{KParameters},
\end{equation}
we obtain that the eigenvalue ${E}^{(1)}_{I}$ is given by
\begin{equation}
\hat{E}^{(1)}_{I}|{\psi'}^{(1)}_{n}\rangle=\pm\frac{1}{2}\sqrt{(\hbar\omega_{0}-k\hbar\omega)^{2}+4\lambda\hat{a}^{k}(\hat{a}^{\dag})^{k}}|{\psi'}^{(1)}_{n}\rangle.
\end{equation}
In this proceeding, we have considered an anti-normal ordering for the multiplication of the creation and annihilation operators, and that the operator $\frac{1}{(\hat{a}^{\dag})^{k}}$ is the inverse operator of $(\hat{a}^{\dag})^{k}$. Therefore, since the eigenvalues of the operator $\hat{a}^{k}(\hat{a}^{\dag})^{k}$ are $\frac{(n+k)!}{n!}$, the energy spectrum ${E}^{(1)}_{I}$ is
\begin{equation}
{E}^{(1)}_{I}=\pm\frac{1}{2}\sqrt{(\hbar\omega_{0}-k\hbar\omega)^{2}+4\lambda\frac{(n+k)!}{n!}}\label{kEn1}, \quad\quad n=0, 1, 2,...
\end{equation}
For the vector ${\phi'}^{(2)}_{n'}$ we can consider the coherent state parameters of equation (\ref{Parameters2}), but with the terms $(\hat{a}^{\dag})^{k}$ and $\hat{a}^{k}$ instead of $(\hat{a}^{\dag})^{2}$ and $\hat{a}^{2}$.
With these elections and a procedure analogous to that made to ${\phi'}^{(1)}_{n'}$, we find that the  energy spectrum ${E}^{(2)}_{I}$ is given by
\begin{equation}
{E}^{(2)}_{I}=\pm\frac{1}{2}\sqrt{(\hbar\omega_{0}-k\hbar\omega)^{2}+4\lambda\frac{n'!}{(n'-k)!}}\label{kEn2}, \quad\quad n'=k, k+1, k+2,...
\end{equation}

With all these results, we are able to construct the eigenfunctions $\Psi'$ of the matrix Hamiltonian $H'_{I}$ if we impose that spinor components $|{\psi'}^{(1)}_{n}\rangle$ and $|{\psi'}^{(2)}_{n'}\rangle$ satisfy the relationship $n'\Rightarrow n+k$. Therefore,
\begin{equation}
\Psi'_{j}=\begin{pmatrix}
|{\psi'}^{(1)}_{n}\rangle\\
|{\psi'}^{(2)}_{n+k}\rangle
\end{pmatrix},\label{kestado2}  \quad\quad n=0, 1, 2,...
\end{equation}
where $|{\psi'}^{(1)}_{n}\rangle$ and $|{\psi'}^{(2)}_{n}\rangle$ are the number states of the one-dimensional harmonic oscillator.

The eigenfunctions $\Psi_{n}$ are obtained from the relationship $\Psi_{n}=D(\xi)\Psi'_{n}$ where $D(\xi)$ is the displacement operator of equation (\ref{Mat}). However, for this case $\Delta=\hbar\omega_{0}-k\hbar\omega$ and $E_{I}$ is the energy eigenvalue of the spinor of equation (\ref{kestado2})
\begin{equation}
E_{I}=\pm\frac{1}{2}\sqrt{(\hbar\omega_{0}-k\hbar\omega)^{2}+4\lambda\frac{(n+k)!}{n!}}\label{kEf}, \quad\quad n=0, 1, 2,...
\end{equation}
Hence, the normalized spinor of the k-photon Jaynes-Cummings model of equation (\ref{KPH}) is given by
\begin{equation}
\Psi_{j}=\begin{pmatrix}
\frac{\sqrt{2}E_{i}}{\sqrt{E_{I}+\Delta}-\sqrt{E_{I}-\Delta}}|{\psi}^{(1)}_{n}\rangle\\\frac{\sqrt{2}E_{I}}{\sqrt{E_{I}+\Delta}+\sqrt{E_{I}-\Delta}}|{\psi}^{(2)}_{n+k}\rangle
\end{pmatrix},\label{estado4}  \quad n=0, 1, 2,...
\end{equation}
with its corresponding energy eigenvalues
\begin{equation}
E_{n}=\hbar\omega\frac{(2n+k)}{2}\pm\frac{1}{2}\sqrt{(\hbar\omega_{0}-k\hbar\omega)^{2}+4\lambda\frac{(n+k)!}{n!}}. \quad\quad n=0, 1, 2,...
\end{equation}
This energy spectrum of the k-photon Jaynes-Cummings model matches perfectly with that presented in reference \cite{Pana}, obtained by means of the Bargmann-Segal representation. Also, it is important to note that for $k=1$, this energy spectrum is adequately reduced to that of the Jaynes-Cummings model. If we set $k=2$, this energy spectrum coincides with that of equation (\ref{energy}), obtained in the previous Section.

\section{Concluding remarks}

We have obtained the energy spectrum and eigenfunctions of the two-photon and k-photon Jaynes-Cummings models. In order to get the exact solution of this problem, we diagonalized the interaction Hamiltonian of these models by introducing a novelty algebraic method. This matrix diagonalization was carried out by means of the $SU(2)$ realization of the Pauli matrices and the tilting transformation, which requires an adequate choice of coherent state parameters. After the diagonalization, we were able to find the energy spectrum of these quantum optic models by chossing a particular order for the multiplication of the creation and annihilation operators. The importance of this $SU(2)$ diagonalization lies in the fact that it is no longer necessary to decouple the equations of the spinor components. The eigenfunctions of each model were obtained by introducing a matrix form of the tilting operator, which in this case is a $2\times 2$ matrix.

It is important to note that this problem can be also solved by using the $su(1,1)$ realization of the creation and annihilation operators $K_0=\frac{1}{2}\left(a^{\dag}a+\frac{1}{2}\right)$, $K_+=\frac{1}{2}a^{\dag^2}$, $K_-=\frac{1}{2}a^2$ and the tilting transformation of this group. However, in this alternative method it is necessary to decouple first the differential equation of the component spinors, as can be seen in reference \cite{Nos}.

\section{Appendix.}

\subsection{ $SU(2)$ Perelomov number coherent states}

The $su(2)$ Lie algebra is spanned by the generators $J_{+}$, $J_{-}$ and $J_{0}$, which satisfy the commutation relations \cite{Vourdas}
\begin{eqnarray}
[J_{0},J_{\pm}]=\pm J_{\pm},\quad\quad [J_{+},J_{-}]=2J_{0}.\label{com2}
\end{eqnarray}
The Casimir operator $J^2$ in this representation is
\begin{equation}
J^{2}=J_0^2+\frac{1}{2}\left(J_+J_-+J_-J_+\right).
\end{equation}
The action of these operators on the Dicke space states (angular momentum states)\\
$\{|j,\mu\rangle, -j\leq\mu\leq j\}$ is
\begin{equation}
J_{+}|j,\mu \rangle=\sqrt{(j-\mu)(j+\mu+1)}|j,\mu+1 \rangle,\label{j+m}
\end{equation}
\begin{equation}
J_{-}|j,\mu \rangle=\sqrt{(j+\mu)(j-\mu+1)}|j,\mu-1 \rangle,\label{j-m}
\end{equation}
\begin{equation}
J_{0}|j,\mu \rangle=\mu|j,\mu \rangle.\label{j0m}
\end{equation}
\begin{equation}
J^2|j,\mu \rangle=j(j+1)|j,\mu \rangle.
\end{equation}
The displacement operator $D(\xi)$ for this Lie algebra is
\begin{equation}
D(\xi)=\exp(\xi J_{+}-\xi^{*}J_{-}),\label{D}
\end{equation}
where $\xi=-\frac{1}{2}\theta e^{-i\varphi}$. By means of Gaussian decomposition, the normal form of this  operator is given by
\begin{equation}
D(\xi)=\exp(\zeta J_{+})\exp(\eta J_{0})\exp(-\zeta^*J_{-})\label{normal2},
\end{equation}
where $\zeta=-\tan(\frac{1}{2}\theta)e^{-i\varphi}$ and $\eta=-2\ln \cos |\xi|=\ln(1+|\zeta|^2)$.

The $SU(2)$ Perelomov coherent states $|\zeta\rangle=D(\xi)|j,-j\rangle$ are given by  \cite{Perellibro,Arecchi}
\begin{equation}
|\zeta\rangle=\sum_{\mu=-j}^{j}\sqrt{\frac{(2j)!}{(j+\mu)!(j-\mu)!}}(1+|\zeta|^{2})^{-j}\zeta^{j+\mu}|j,\mu\rangle,\label{PCN2}
\end{equation}
The $SU(2)$ Perelomov number coherent state $|\zeta,j,\mu\rangle$ is defined as the action of the displacement operator $D(\xi)$ onto an arbitrary
excited state $|j,\mu\rangle$
\begin{eqnarray}
|\zeta,j,\mu\rangle &=&\sum_{s=0}^{j-\mu+n}\frac{\zeta^{s}}{s!}\sum_{n=0}^{\mu+j}\frac{(-\zeta^*)^{n}}{n!}e^{\eta(\mu-n)}
\frac{\Gamma(j-\mu+n+1)}{\Gamma(j+\mu-n+1)}\nonumber\\ &&\times\left[\frac{\Gamma(j+\mu+1)\Gamma(j+\mu-n+s+1)}{\Gamma(j-\mu+1)\Gamma(j-\mu+n-s+1)}\right]^{\frac{1}{2}}|j,\mu-n+s\rangle.\label{PNCS2}
\end{eqnarray}

The tilting transformation of the $su(2)$ Lie algebra generators are computed by using of the displacement operator $D(\xi)$ are give by
\begin{equation}
D^{\dag}(\xi)J_{+}D(\xi)=-\frac{\xi^{*}}{|\xi|}\delta J_{0}+\epsilon\left(J_{+}+\frac{\xi^{*}}{\xi}J_{-}\right)+J_{+}\label{DJ+},
\end{equation}
\begin{equation}
D^{\dag}(\xi)J_{-}D(\xi)=-\frac{\xi}{|\xi|}\delta J_{0}+\epsilon\left(J_{-}+\frac{\xi}{\xi^{*}}J_{+}\right)+J_{-}\label{DJ-},
\end{equation}
\begin{equation}
D^{\dag}(\xi)J_{0}D(\xi)=(2\epsilon+1)J_{0}+\frac{\delta\xi}{2|\xi|}J_{+}+\frac{\delta\xi^{*}}{2|\xi|}J_{-}\label{DJ0},
\end{equation}
where $\delta=\sin(2|\xi|)$ and $\epsilon=\frac{1}{2}\left[\cos(2|\xi|)-1\right]$.

A particular realization of the $su(2)$ Lie algebra is given in terms of the Pauli matrices
\begin{equation}
\sigma_{0}=\begin{pmatrix} 1&0\\ 0&-1\\\end{pmatrix}, \quad\quad
 \sigma_{+}=\begin{pmatrix} 0&1\\ 0&0\\\end{pmatrix}, \quad\quad
 \sigma_{-}=\begin{pmatrix} 0&0\\ 1&0\\\end{pmatrix}.
\end{equation}
For this realization, the operators $J_0$, $J_+$ and $J_-$ are defined as
\begin{equation}
J_{0}=\frac{\sigma_{0}}{2}, \quad\quad J_{+}=\sigma_{+}, \quad\quad J_{-}=\sigma_{-}.\label{real}
\end{equation}


\begin{thebibliography} {100}

\bibitem{Jay} E.T. Jaynes, and F.W. Cummings, Proc. IEEE 51, 89 (1963).

\bibitem{Dod} V.V. Dodonov, and V.I. Man'ko (eds), Theory of Nonclassical States of Light, Taylor and Francis, London, 2003.

\bibitem{shore} B.W. Shore, and P.L. Knight, J. Mod. Opt. 40, 1195 (1993).

\bibitem{Haroche} S. Haroche, J.M. Raimond, Exploring the Quantum: Atoms, Cavities and Photons, Oxford
University Press, Oxford, 2007.

\bibitem{Narozhny} J.H. Eberly, N.B. Narozhny, and J.J. Sanchez-Mondragon, Phys. Rev. Lett. 44, 1323 (1980).

\bibitem{Kuklinski} J.R. Kuklinski, and J. Madajczyk, Phys. Rev. A 37, 3175 (1988).

\bibitem{Goy} P. Goy, J.M. Raimond, M. Gross, and S. Haroche, Phys. Rev. Lett. 50, 1903 (1983).

\bibitem{Brune} M. Brune et al., Phys. Rev. Lett. 76, 1800 (1996).

\bibitem{Guerlin} C. Guerlin et al., Nature 448, 889 (2007).

\bibitem{Toyama} Y. Nogami, and F.M. Toyama, Can. J. Phys. 74, 114 (1996).

\bibitem{Nogami} F.M. Toyama, Y. Nogami, and F.A.B. Coutinho, J. Phys. A: Math. Gen. 30, 2585 (1997).

\bibitem{Rados} R. Szmytkowski, and M. Gruchowski, J. Phys. A: Math. Gen. 34, 4991 (2001).

\bibitem{Nos} D. Ojeda-Guill\'en, R.D. Mota, and V.D. Granados, J. Math. Phys. 57, 062104 (2016).

\bibitem{Bermudez} A. Bermudez, M.A. Martin-Delgado, and E. Solano, Phys. Rev. A 76, 041801 (2007).

\bibitem{Chiral} A. Bermudez, M. A. Martin-Delgado, and A. Luis, Phys Rev. A 77, 063815 (2008).

\bibitem{Koch} E.A. Kochetov, J. Phys. A 20, 2433 (1987).

\bibitem{Buze} V. Buzek, Czech. J. Phys. B 39, 757 (1989).

\bibitem{Gou} S.C. Gou, Phys. Rev. A 48, 3233 (1993).

\bibitem{JCAJC} E. Chore$\tilde{n}$o, D. Ojeda-Guill\'en, M. Salazar-Ram\'irez, and V.D. Granados, Ann. Phys. 387, 121 (2017).

\bibitem{Lam} L. Lamata et al., New J. Phys. 13, 095003 (2011).

\bibitem{Ret} A. Retzker, E. Solano, and B. Reznik, Phys. Rev. A 75, 022312 (2007).

\bibitem{Kop} W. Kopylov et al., Phys. Rev. A 92, 063832 (2015).

\bibitem{Sun} C. Sun, and N. Sinitsyn, Phys. Rev. A, 033808 (2016).

\bibitem{Cha} M. Chaichian, D. Ellinas, and P. Kulish, Phys. Rev. Lett. 65, 980 (1990).

\bibitem{Buck80} B. Buck, and C.V. Sukumar, Phys. Lett. 81A, 132 (1981).

\bibitem{Gerry} C.C. Gerry, Phys. Rev. A 37, 2683 (1988).

\bibitem{Joshi}  A. Joshi, and R.R. Puri, Phys. Rev. A 45, 5056 (1992).

\bibitem{Bartzis} V. Bartzis and N. Nayak, J. Opt. Soc. Am. B 8, 1779 (1991).

\bibitem{Zhou} P. Zhou, and J.S. Peng, Phys. Rev. A 44, 3331 (1991).

\bibitem{Tahira} T. Nasreen, and M.S.K. Razmi, J. Opt. Soc. Am. B 10, 1292 (1993).

\bibitem{Ng} K.M. Ng, C.F. Lo, and K.L. Liu, Eur. Phys. J. D 6, 119 (1999).

\bibitem{Sukumar81} C.V. Sukumar, and B. Buck, Phys. Lett. 83A, 211 (1981).

\bibitem{Shumovsky} A.S. Shumovsky, F.L. Kien, and E.I. Aliskenderov, Phys. Lett. A 124, 351 (1987).

\bibitem{Vogel} W. Vogel, and D.G. Welsch, Phys. Rev. A 40, 7113 (1989).

\bibitem{Baghshahi} H.R. Baghshahi, M.K. Tavassoly, and A. Behjat, Chin. Phys. B 23, 074203 (2014).

\bibitem{kocr} R. Koc, H. Tutunculer, M. Koca, and E. Olgar, Ann. Phys. 319, 333 (2005).

\bibitem{Pana} H. Panahi, and S.A. Rad, Int. J. Theor. Phys. 52, 4068 (2013).

\bibitem{gerryberry} C.C. Gerry, Phys. Rev. A 39, 3204 (1989).

\bibitem{Nos1} D. Ojeda-Guill\'en, R.D. Mota, and V.D. Granados, J. Math. Phys. 55, 042109 (2014).

\bibitem{Nos2} D. Ojeda-Guill\'en, R.D. Mota, and V.D. Granados, Commun. Theor. Phys. 64 34 (2015).

\bibitem{Vourdas}  A. Vourdas, Phys. Rev. A 41, 1653 (1990).

\bibitem{Perellibro} A.M. Perelomov, Generalized Coherent States and Their Applications, Springer, Berlin, 1986.

\bibitem{Arecchi}  F.T. Arecchi, E. Courtens, R. Gilmore and H. Thomas, Phys. Rev. A 6, 2211 (1972).

\end{thebibliography}
\end{document}